\documentclass[a4paper,12pt]{article}
\usepackage{amsfonts}
\usepackage{latexsym}
\usepackage{amsmath}
\usepackage{amssymb}
\usepackage{amssymb}
\usepackage{slashed}
\usepackage{upgreek}
\usepackage{mathrsfs}
\usepackage{bbm}

\usepackage{relsize}
\usepackage{graphicx}

\newcommand{\newsection}{    
\setcounter{equation}{0}\section}
\def\appendix#1{\addtocounter{section}{1}\setcounter{equation}{0}
\renewcommand{\thesection}{\Alph{section}}
\section*{Appendix \thesection\protect\indent \parbox[t]{11.15cm}{#1}}
\addcontentsline{toc}{section}{Appendix \thesection\ \ \ #1}}

\def\bbe{{\bf{e}}}
\font\mybb=msbm10 at 11pt

\def\bb#1{\hbox{\mybb#1}}

\def\bR {\bb{R}}

\def\gX{\Gamma\mkern-4.0mu X}
\def\gom{\Gamma\mkern-4.0mu \omega}

\def\gF{\Gamma\mkern-4.0mu F}

\def\gH{\Gamma\mkern-4.0mu H}
\def\gG{\Gamma\mkern-4.0mu G}

\def\sX{\slashed {X}}
\def\sgX{\slashed {\gX}}

\def\sgF{\slashed {\gF}}

\def\sgH{\slashed{\gH}}
\def\sgG{\slashed{\gG}}

\newcommand{\bea}{\begin{eqnarray}}
\newcommand{\eea}{\end{eqnarray}}

\begin{document}
\begin{titlepage}
\begin{center}
\vspace*{-1.0cm}
\hfill DMUS--MP--16/02 \\

\vspace{2.0cm} {\Large \bf AdS$_5$ Backgrounds with 24 Supersymmetries} \\[.2cm]

\vskip 2cm
S. W. Beck$^1$,  J. B.  Gutowski$^2$ and G. Papadopoulos$^1$
\\
\vskip .6cm

\begin{small}
$^1$\textit{  Department of Mathematics, King's College London
\\
Strand, London WC2R 2LS, UK.\\
E-mail: george.papadopoulos@kcl.ac.uk}
\end{small}\\*[.6cm]

\begin{small}
$^2$\textit{Department of Mathematics,
University of Surrey \\
Guildford, GU2 7XH, UK \\
Email: j.gutowski@surrey.ac.uk}
\end{small}\\*[.6cm]

\end{center}

\vskip 3.5 cm
\begin{abstract}

We prove a non-existence theorem for smooth $AdS_5$ solutions with  connected, compact without boundary internal space that preserve strictly 24 supersymmetries.
In particular, we show that   $D=11$ supergravity does not admit such solutions, and that all such solutions of IIB supergravity
are locally
isometric to the $AdS_5 \times S^5$  maximally supersymmetric background.
Furthermore, we prove that (massive) IIA supergravity also
does not admit such solutions, provided that the homogeneity
conjecture for massive IIA supergravity is valid.
In the context of AdS/CFT these results imply that if strictly  ${\cal N}=3$  superconformal theories in 4-dimensions exist, their gravitational dual backgrounds  are either singular
or their internal spaces are not compact.

\end{abstract}

\end{titlepage}


\section{Introduction}
Warped $AdS_5$ backgrounds of 10- and 11-dimensional supergravity theories are of particular interest within the
AdS/CFT correspondence as they are dual to 4-dimensional superconformal theories, see \cite{maldacena} for a review.  The most celebrated example
of such correspondence is the statement that  IIB superstring theory on the maximally supersymmetric $AdS_5\times S^5$ background is dual to ${\cal N}=4$ supersymmetric gauge theory in four dimensions \cite{maldacena0}. AdS spaces have been used in supergravity compactifications, for a review see \cite{duff}.

To establish such  dualities to other 4-dimensional superconformal theories with  less than maximal  supersymmetry requires the
construction of $AdS_5$ supergravity backgrounds preserving less than 32 supersymmetries.  Recently it has been shown in \cite{mads, iibads, iiaads} that
$AdS_5$ backgrounds preserve 8, 16, 24 or 32 supersymmetries in type II 10-dimensional supergravities and in 11-dimensional supergravity\footnote{There
are no supersymmetric  $AdS_5$ backgrounds in either heterotic or type I supergravities.}.  The  maximally supersymmetric $AdS_5$ backgrounds have been
classified in \cite{maxsusy} where it has been  shown that no such backgrounds exist in either  11-dimensional  or (massive) IIA supergravities, and all maximally supersymmetric $AdS_5$
backgrounds in IIB supergravity are locally isometric to the previously known  $AdS_5\times S^5$ solution of the theory, see \cite{schwarz} and reference within.
To our knowledge there is no classification of $AdS_5$ backgrounds preserving 16 or 24 supersymmetries.  The geometry of $AdS_5$ solutions preserving 8 supersymmetries
 has been  investigated
in \cite{dario1, dario2}, after assuming that the fields are invariant under the $\mathfrak{so}(4,2)$ symmetry of $AdS_5$ together with some
   additional restrictions\footnote{Typically it is assumed that the Killing spinors factorize as  Killing spinors on AdS and  Killing spinors
  along the internal space. A factorization of this type has been investigated in \cite{mads, iibads, iiaads} and it was found that it imposes
  more restrictions on the backgrounds than those required for invariance under the isometries of AdS. Therefore the generality
  of the factorization approaches must be re-investigated on a case by case basis.} on the form  of Killing spinors. Moreover, many $AdS_5$ solutions have been found, see for example \cite{spence}-\cite{passias2}. In
 \cite{mads, iibads, iiaads} a different approach to investigate the geometry of AdS backgrounds was proposed, which was based on
    earlier work on black hole near horizon geometries \cite{adshor}  which has the advantage that all additional  restrictions are  removed and the only assumption that remains
    is the requirement for  the fields to be invariant under the $\mathfrak{so}(4,2)$ symmetry of $AdS_5$.

In this paper, we shall demonstrate, under the assumptions we describe in detail below, that there are no $AdS_5$ solutions in 11-dimensional and (massive) IIA supergravities  that preserve 24 supersymmetries.    Furthermore we shall show that all $AdS_5$ solutions of IIB
supersgravity that preserve 24 supersymmetries are locally isometric to the maximally supersymmetric  $AdS_5\times S^5$ background.

One application of our results is in AdS/CFT and in particular on the existence of gravitational duals for strictly ${\cal N}=3$ superconformal theories in four dimensions.
It is known for sometime that the field content and component actions of  ${\cal N}=3$ and ${\cal N}=4$ superconformal theories with rigid supersymmetry
are the same.  As a result  ${\cal N}=3$ superconformal symmetry classically enhances to ${\cal N}=4$. Quantum mechanically, the picture
is more involved as the quantization of these theories with manifest ${\cal N}=3$ and ${\cal N}=4$ supersymmetry will require the use techniques like
harmonic superspace \cite{galperin, delduc},  and these are different  for these two theories.  Nevertheless the  interpretation of the equivalence of the classical actions is that perturbatively the two quantum theories are indistinguishable\footnote{ In fact it may be possible to prove this by demonstrating via Ward identity techniques like those in \cite{white} that
${\cal N}=3$ superconformal symmetry quantum mechanically always enhances  to ${\cal N}=4$. We would like to thank Paul Howe for suggesting this.}. Therefore if a theory exists with strictly ${\cal N}=3$ superconformal symmetry,
it must be intrinsically  non-perturbative. The properties of such ${\cal N}=3$ superconformal theories have been investigated in \cite{aharony2} and an F-theory construction
for such a  theory has been proposed in \cite{garcia}.
In this context our results imply that, unlike the ${\cal N}=4$ superconformal theories, there are no smooth gravitational duals,
with compact without boundary internal spaces, for strictly ${\cal N}=3$ superconformal theories in four dimensions.

The proof of the above result utilizes  the near horizon approach of \cite{mads, iibads, iiaads} for solving the Killing spinor equations (KSEs) of supergravity theories for AdS backgrounds as well as a technique developed for the proof
of the homogeneity conjecture in \cite{homogen}. Furthermore, to prove our  results we make   certain smoothness and global assumptions. In particular apart from  implementing the $\mathfrak{so}(4,2)$ symmetry on the fields, we also assume that the warped $AdS_5\times_w M^{D-5}$ backgrounds, for $D=10$ or $D=11$,   satisfy the following conditions\footnote{We also assume the validity of the homogeneity conjecture  for massive IIA supergravity. This has not been proven as yet but it is expected to hold.}: (i) All the fields are smooth, and (ii) the internal space $M^{D-5}$ is  connected\footnote{In fact $M^{D-5}$ is required to be path connected but all manifolds are path connected
 if they are connected since they are locally path connected. From now on, we shall assume that $M^{D-5}$ is always connected.} and compact without boundary.
Both these additional restrictions, apart from the connectness of $M^{D-5}$,
can   be replaced with the assertion that the data are such that the Hopf maximum principle applies.
 These assumptions are essential as otherwise there are
for example $AdS_5$ backgrounds in 11-dimensions which preserve more than 16 supersymmetries, see also section 2.

This paper is organized as follows.  In section 2, we prove the non-existence of $AdS_5$ backgrounds preserving 24 supersymmetries
in 11-dimensional supergravity, and in section 3, we demonstrate the same result for both standard and massive IIA supergravities.  In section 4,
we show  that the $AdS_5$ backgrounds that preserve 24 supersymmetries
in  IIB supergravity are locally isometric to the maximally supersymmetric background $AdS_5\times S^5$. In section 5, we give our conclusions and
explore an application to AdS/CFT. Furthermore, in appendix A we briefly summarize some of our conventions, and in appendix B for completeness we present
a technique we use to derive our results  which has been adapted from the proof of the homogeneity conjecture.

\newsection{ $AdS_5 \times _w M^6$ Solutions in D=11}

We begin by briefly summarizing the general structure of warped $AdS_5$ solutions in 11-dimensional supergravity, as determined
in \cite{mads}, whose conventions we shall follow throughout this section.
Then we shall present the proof that there are no such solutions preserving
24 supersymmetries. The metric and 4-form are given by
\begin{eqnarray}
ds^2 &=& 2 du (dr + r h )+ A^2 (dz^2+ e^{2z/\ell} \sum_{a=1}^2 (dx^a)^2) +ds^2(M^6)~,
\cr
F &=&   X~,
\label{mhmadsn}
\end{eqnarray}
where  we have written the solution as a near-horizon geometry \cite{adshor},
with
\bea
h = -{2 \over \ell}dz -2 A^{-1} dA~,
\eea
$(u,r, z, x^1, x^2)$ are the coordinates of the $AdS_5$ space,  $A$ is the warp factor which  is a function on $M^6$, and $X$ is a closed 4-form on $M^6$.
$A$ and $X$ depend only on the coordinates of $M^6$, $\ell$ is the radius of $AdS_5$.

The 11-dimensional Einstein equation implies that
\bea
\label{lapein}
D^k \partial_k \log A=-{4\over \ell^2} A^{-2}- 5 \partial^k \log A\, \partial_k \log A+{1\over 144}  X^2~,
\eea
where $D$ is the Levi-Civita connection on $M^6$.
The remaining components of the Einstein and gauge field equations
are listed in \cite{mads}, however we shall only require ({\ref{lapein}})
for the analysis of the $N=24$ solutions.
In particular, ({\ref{lapein}}) implies that $A$ is everywhere non-vanishing
on $M^6$, on assuming that $M^6$ is connected and all fields are smooth.

We adopt the following frame conventions; $\bbe^i$ is
an orthonormal frame for $M^6$, and
\begin{eqnarray}
\label{orthfr}
\bbe^+ = du~, \qquad \bbe^- = dr+rh~, \qquad
\bbe^z=A dz~, \qquad \bbe^a = A e^{z/\ell} dx^a~.
\eea
We use this frame in the investigation of KSEs below.

\subsection{The Killing spinors}

The Killing spinors of $AdS_5$ backgrounds are given by
\begin{eqnarray}
\epsilon&=&\sigma_+-\ell^{-1} \sum_{a=1}^{2} x^a\Gamma_{az} \tau_++ e^{-{z\over\ell}} \tau_++\sigma_-+e^{{z\over\ell}}(\tau_--\ell^{-1} \sum_{a=1}^{2} x^a \Gamma_{az} \sigma_-)
\cr
&&-\ell^{-1} u A^{-1} \Gamma_{+z} \sigma_--\ell^{-1} r A^{-1}e^{-{z\over\ell}} \Gamma_{-z}\tau_+~,
\label{kkk}
\end{eqnarray}
where we have used the light-cone projections
\begin{eqnarray}
\Gamma_\pm \sigma_\pm =0~,~~~\Gamma_\pm\tau_\pm=0~,
\label{dec}
\end{eqnarray}
and $\sigma_\pm$ and $\tau_\pm$ are 16-component spinors that depend only on the coordinates of $M^6$.
We do not assume that the Killing spinors factorize as Killing spinors on $AdS_5$ and Killing spinors on $M^6$.

The remaining independent Killing spinor equations (KSEs) are:
\bea
D^{(\pm)}_i \sigma_\pm=0~,~~~D^{(\pm)}_i \tau_\pm=0~,
\label{kseadsk1}
\eea
and
\bea
{\Xi}^{(\pm)} \sigma_\pm=0~,~~~{\Xi}^{(\mp)} \tau_\pm=0~,
\label{kseadsk2}
\eea
where
\bea
D^{(\pm)}_i&=&D_i \pm {1\over 2} \partial_i \log A-{1\over 288} \sgX_i
+{1\over 36} \sX_i~,
\cr
\Xi^{(\pm)}&=&-{1\over2}\Gamma_z \Gamma^i \partial_i \log A\mp {1\over 2\ell}  A^{-1} +{1\over 288}\Gamma_z \sX~.
\eea
In particular   algebraic KSEs ({\ref{kseadsk2}}) imply that $\sigma_+$ and $\tau_+$ cannot be linearly dependent. For our Clifford algebra conventions see also appendix A.

\subsection{Counting the Killing Spinors}

In order to count the number of supersymmetries, note that if
$\sigma_+$ is a solution of the $\sigma_+$ KSEs, then so is $\Gamma_{12} \sigma_+$.
Furthermore, $\tau_+=\Gamma_z \Gamma_1 \sigma_+$ and $\tau_+ = \Gamma_z \Gamma_2 \sigma_+$
are solutions to the $\tau_+$  KSEs. The spinors $\sigma_+, \Gamma_{12} \sigma_+, \Gamma_z \Gamma_1 \sigma_+, \Gamma_z \Gamma_2 \sigma_+$ are linearly independent.
The positive chirality spinors also generate negative chirality spinors
$\sigma_-$, $\tau_-$ which satisfy the appropriate KSEs. This is
because if  $\sigma_+, \tau_+$ is a solution, then so is
\bea
\sigma_-=A \Gamma_- \Gamma_z \sigma_+~,~~~\tau_-=A \Gamma_- \Gamma_z \tau_+~,~~~
\eea
and also conversely, if $\sigma_-, \tau_-$ is a solution, then so is
\bea
\sigma_+=A^{-1} \Gamma_+ \Gamma_z \sigma_-~,~~~\tau_+=A^{-1} \Gamma_+ \Gamma_z \tau_-~.~~~
\label{pmrel}
\eea
So for a generic $AdS_5 \times_w M^6$ solution, all of the Killing
spinors are generated by the $\sigma_+$ spinors, each of which
gives rise to 8 linearly independent spinors via the mechanism described here. The solutions therefore preserve $8k$ supersymmetries, where $k$ is equal
to the number of $\sigma_+$ spinors.

\subsection{Non-existence of $N=24$ $AdS_5$ solutions in D=11}

To consider the $AdS_5$ solutions preserving 24 supersymmetries, we begin by setting
\bea
 \Lambda = \sigma_+ + \tau_+
\eea
and defining
\bea
\label{modiso}
W_i = A \langle \Lambda, \Gamma_{z12} \Gamma_i \Lambda \rangle \ .
\eea
Then ({\ref{kseadsk1}}) implies that
\bea
D_{(i} W_{j)}=0
\eea
so $W$ is an isometry of $M^6$. In addition, the algebraic conditions ({\ref{kseadsk2}})
imply that
\bea
\label{extraglob1}
{1 \over 288} \langle \Lambda , \sgX_i   \Lambda \rangle -{1 \over 2} \parallel \Lambda \parallel^2
A^{-1} D_i A - \ell^{-1} A^{-1} \langle \tau_+ , \Gamma_i \Gamma_z \sigma_+ \rangle =0 \ .
\eea
Also, ({\ref{kseadsk1}}) implies that
\bea
\label{extraglob2}
D_i \parallel \Lambda \parallel^2 = - \parallel \Lambda \parallel^2
A^{-1} D_i A  +{1 \over 144} \langle \Lambda , \sgX_i \Lambda \rangle \ .
\eea
Combining ({\ref{extraglob1}}), and ({\ref{extraglob2}}) we have
\bea
\label{extraglob3}
D_i \parallel \Lambda \parallel^2 - 2\ell^{-1} A^{-1}\langle \tau_+ , \Gamma_i \Gamma_z \sigma_+ \rangle =0 \ .
\eea
In addition ({\ref{kseadsk1}}) implies that
\bea
D^i \bigg(A \langle \tau_+ , \Gamma_i \Gamma_z \sigma_+ \rangle \bigg) =0 \ .
\eea
Hence, on taking the divergence of ({\ref{extraglob3}}), we find
\bea
D^i D_i  \parallel \Lambda \parallel^2 +2 A^{-1} D^i A D_i  \parallel \Lambda \parallel^2 =0 \ .
\eea
A maximum principle argument then implies that $ \parallel \Lambda \parallel^2 $ is constant.
Substituting these conditions back into ({\ref{extraglob2}}), we find the condition
\bea
\label{newglob1}
i_W H = 6 \parallel \Lambda \parallel^2 dA~,
\eea
where
\bea
H = \star_6 X~,
\eea
and $\star_6$ denotes the Hodge dual on $M^6$.

To prove a non-existence theorem for $N=24$ solutions, we
consider spinors of the type
\bea
\Lambda = \sigma_+ + \tau_+ \ .
\eea
For a $N=24$ solution, there are 12 linearly independent spinors
of this type, because of the algebraic conditions ({\ref{kseadsk2}}).
Next, consider the condition ({\ref{newglob1}}). This implies that
\bea
i_W dA=0~,
\eea
where $W$ is the isometry generated by $\Lambda$ as defined in ({\ref{modiso}}).

A straightforward  modification of the reasoning used in
\cite{homogen}, which we describe in Appendix B, implies that for $N=24$ solutions, the vector fields
dual to the 1-form bilinears $W$ generated by the $\Lambda$ spinors span the tangent space of $M^6$.
Then the condition $i_W dA=0$ implies that $A$ is constant,
and furthermore, ({\ref{newglob1}}) implies that $i_W H=0$, which
also implies that $H=0$, and so $X=0$.

However, the Einstein equation ({\ref{lapein}}) admits no $AdS_5$ solutions for which
$dA=0$ and $X=0$, so there can be no $N=24$ $AdS_5$ solutions.

We should remark that the two assumptions we have made on the fields to derive this result are essential. This is because  any AdS$_{d+1}$ background can locally be written as a warped product
$ds^2(AdS_{d+1})= dy^2 + A^2(y) ds^2(AdS_{d})$ for some function $A$ which has been determined in   \cite{gutpap}. For $d=2$, this has  previously been established in \cite{strominger}.
As a result the maximally supersymmetric $AdS_7\times S^4$ solution of 11-dimensional supergravity can be seen as a warped $AdS_5$ background. This appears to be
 a contradiction to our result. However, the internal space $M^6$ in this case is non-compact and so it does not satisfy the two assumptions we have made.

\newsection{ $AdS_5 \times _w M^5$ solutions in (massive) IIA supergravity}

As in the 11-dimensional supergravity investigated in the previous sections, there are no $N=24$ AdS5 backgrounds in (massive) IIA
supergravity. We shall use the formalism and follow the conventions of \cite{iiaads} in the analysis that follows.
Imposing invariance of the background under the symmetries of $AdS_5$ all the fluxes are magnetic, ie their components along $AdS_5$ vanish.
 In particular the most general $AdS_5$ background is
\begin{eqnarray}
&&ds^2 = 2 du (dr + r h)+ A^2 \big(dz^2+ e^{2z/\ell} \sum_{a=1}^{2}(dx^a)^2\big) +ds^2(M^{5})~,
\cr
&&G = G~,~~~
H=H~,~~F=F~,~~\Phi=\Phi~,~~S=S~,~~h = -{2 \over \ell}dz -2 A^{-1} dA~,
\end{eqnarray}
where we have denoted the 10-dimensional fluxes and their components along $M^5$ with the same symbol,   $A$ is the warp factor, $\Phi$ is the dilaton and   $S$ is the cosmological constant dressed with the dilaton.  $A$, $S$ and $\Phi$  are functions of $M^5$, while  $G$ , $H$  and $F$ are the 4-form, 3-form and a 2-form fluxes, respectively, which have support only on $M^5$. The coordinates of $AdS_5$ are $(u,r,z, x^a)$ and we introduce the frame $(\bbe^+, \bbe^-, \bbe^z, \bbe^a)$ as in (\ref{orthfr}).

The fields satisfy a number of field equations and Bianchi identities which can be found in \cite{iiaads}. Those relevant
for the analysis that follows are the field equation for the dilaton
and the  field equation for $G$
\begin{align}
 D^2 \Phi &= -5 A^{-1} \partial^i A \partial_i \Phi + 2 ( d\Phi )^2 + \frac{5}{4} S^2 + \frac{3}{8} F^2 - \frac{1}{12} H^2 + \frac{1}{96} G^2~,
\\
 \nabla^\ell G_{i j k \ell} &= -5 A^{-1} \partial^\ell A G_{i j k \ell} + \partial^\ell \Phi G_{i j k \ell}~,
\label{iiafeq}
\end{align}
respectively, and the Einstein equations  both along   $AdS_5$ and $M^5$
\begin{equation}
 D^2 \ln A = -4 \ell^{-2} A^{-2} - 5 A^{-2} ( dA )^2 + 2 A^{-1} \partial_i A \partial^i \Phi + \frac{1}{96} G^2 + \frac{1}{4} S^2 + \frac{1}{8} F^2 ,
 \label{warpfeq}
\end{equation}
\begin{align}
\label{ein5}
 R^{(5)}_{i j} &= 5 \nabla_i \nabla_j \ln A + 5 A^{-2} \partial_i A \partial_j A + \frac{1}{12} G^2_{i j} - \frac{1}{96} G^2 \delta_{i j}
 \\ \nonumber
 & \qquad\qquad - \frac{1}{4} S^2 \delta_{i j} + \frac{1}{4} H^2_{i j} + \frac{1}{2} F^2_{i j} - \frac{1}{8} F^2 \delta_{i j} - 2 \nabla_i \nabla_j \Phi~,
\end{align}
respectively, where $D$ is the Levi-Civita connection of $M^5$ and $ R^{(5)}_{i j}$ is the Ricci tensor of $M^{5}$. The former is seen as
the field equation for the warp factor $A$.

\subsection{Killing spinor equations}

The killing spinors of IIA $AdS_5$ backgrounds are given as in (\ref{kkk})
where now $\sigma_\pm$ and $\tau_\pm$ are 16-component spinors that depend only on the coordinates of $M^5$.  The
remaining independent conditions  are the gravitino KSEs
\begin{eqnarray}
 \nabla^{( \pm )}_i\sigma_\pm&=&0~,~~~ \nabla^{( \pm )}_i\tau_\pm=0~,~~~
\label{kseadsk1iia}
\end{eqnarray}
the dilatino  KSEs
\begin{eqnarray}
 \mathcal{A}^{(\pm)}\sigma_\pm&=&0~,~~~ \mathcal{A}^{(\pm)}\tau_\pm=0~,
 \label{kseadsk2iia}
\end{eqnarray}
and the algebraic  KSEs
\begin{eqnarray}
  \Xi_\pm \sigma_\pm &=& 0~, \qquad \Xi_\pm \tau_\pm = \mp \ell^{-1} \tau_\pm~,
\label{kseadsk3iia}
\end{eqnarray}
where
\begin{eqnarray}
\nabla^{( \pm )}_i&=& D_i + \Psi^{( \pm )}_i~,
\cr
\mathcal{A}^{(\pm)} &=&\slashed{\partial} \Phi + \frac{1}{12} \slashed{H} \Gamma_{11} + \frac{5}{4} S + \frac{3}{8} \slashed{F} \Gamma_{11} + \frac{1}{96} \slashed{G}~,
\cr
\Xi_\pm &=& \mp \frac{1}{2 \ell} + \frac{1}{2} \slashed{\partial} A \Gamma_z - \frac{1}{8} A S \Gamma_z - \frac{1}{16} A \slashed{F} \Gamma_z \Gamma_{11} - \frac{1}{192} A \slashed{G} \Gamma_z~,
\end{eqnarray}
and where $D$ is the spin connection on $M^5$ and
\begin{equation}
 \Psi^{( \pm )}_i = \pm \frac{1}{2 A} \partial_i A + \frac{1}{8} \slashed{H}_i \Gamma_{11} + \frac{1}{8} S \Gamma_i + \frac{1}{16} \slashed{F} \Gamma_i \Gamma_{11} + \frac{1}{192} \slashed{G} \Gamma_i~,
\end{equation}
see  appendix A for our Clifford algebra conventions.
The counting of supersymmetries is exactly the same as in the D=11 supergravity described in the previous sections.

\subsection{$N=24$ $AdS_5$ solutions in (massive) IIA supergravity}

Before we proceed with the analysis, the homogeneity conjecture\footnote{Strictly speaking the homogeneity conjecture has not been proven for massive IIA supergravity, but it is expected to hold.}    \cite{homogen} together with the results  \cite{gran} on the classification of (massive) IIA backgrounds imply that both $\Phi$ and $S$ are constant functions over the whole
spacetime which we shall assume from now on.
Next let us set
\bea
 \Lambda = \sigma_+ + \tau_+~,
\eea
and define
\bea
\label{modisoiia}
W_i = A \langle \Lambda, \Gamma_{zxy} \Gamma_i \Lambda \rangle \ .
\eea
Then ({\ref{kseadsk1iia}}) implies that
\bea
D_{(i} W_{j)}=0~,
\eea
so $W$ is an isometry of $M^5$.

After some straightforward computation using the gravitino KSEs, one finds
\bea
D_i\parallel \Lambda \parallel^2=- A^{-1} \partial_i A \parallel \Lambda \parallel^2-{1\over4} S\langle \Lambda, \Gamma_i \Lambda\rangle-{1\over8}
\langle \Lambda, \sgF_i \Gamma_{11} \Lambda\rangle-{1\over 96} \langle \Lambda, \sgG_i  \Lambda\rangle
\label{iia1}
\eea
On the other hand ({\ref{kseadsk3iia}}) gives
\bea
(\slashed{\partial} A \Gamma_z-{1\over4} AS \Gamma_z-{1\over8} A \slashed{F} \Gamma_z \Gamma_{11}-{1\over 96} A \slashed{G} \Gamma_z)\Lambda=-\ell^{-1} \tau_++ \ell^{-1} \sigma_+~.
\eea
Using this, (\ref{iia1}) can be written as
\bea
D_i\parallel \Lambda \parallel^2=2 \ell^{-1} A^{-1} \langle \tau_+, \Gamma_i \Gamma_z \sigma_+\rangle~.
\label{iia2}
\eea
Furthermore using ({\ref{kseadsk1iia}}), one can show that
\bea
D^i (A \langle \tau_+, \Gamma_i \Gamma_z \sigma_+\rangle)=0~.
\eea
Taking the covariant derivative of (\ref{iia2}) and using the above equation, one finds that
\bea
D^i D_i  \parallel \Lambda \parallel^2 +2 A^{-1} D^i A D_i  \parallel \Lambda \parallel^2 =0 \ .
\eea
This in turn implies after using the maximum principle that $\parallel \Lambda \parallel^2$ is constant.

Using the constancy of $\parallel \Lambda \parallel^2$,  (\ref{iia1}) and (\ref{iia2}) imply that
\bea
- A^{-1} \partial_i A \parallel \Lambda \parallel^2-{1\over4} S\langle \Lambda, \Gamma_i \Lambda\rangle-{1\over8}
\langle \Lambda, \sgF_i \Gamma_{11} \Lambda\rangle-{1\over 96} \langle \Lambda, \sgG_i  \Lambda\rangle=0~,
\label{iia3}
\eea
and
\bea
\langle \tau_+, \Gamma_i \Gamma_z \sigma_+\rangle=0~.
\label{iia4}
\eea
Next taking the difference of the two identities below
\bea
\langle \tau_+, \Xi_+ \sigma_+\rangle=0~,~~~\langle \sigma_+, (\Xi_++\ell^{-1} \tau_+\rangle=0~,
\eea
and upon using (\ref{iia4}), we find
\bea
\langle \tau_+, \sigma_+\rangle=0~,
\label{iia5}
\eea
ie $\tau_+$ and $\sigma_+$ are orthogonal.

To continue, multiply $\Xi_+\Lambda=-\ell^{-1} \tau_+$ with $\Gamma_{xy}$, and using the fact $\Gamma_{xy}\tau_+$ is again a type $\tau_+$ Killing spinor, and the
equation above, one obtains that
\bea
W^i \partial_i A=0~.
\label{iia6}
\eea
As straightforward modification of the argument used in \cite{homogen} to prove the homogeneity conjecture, see also appendix B, one can show that the vector fields $W$ span the tangent spaces  of $M^5$.
As a result, the above equation implies that $A$ is constant.

Next using the dilatino KSE (\ref{kseadsk2iia}) to eliminate the $G$-dependent term in (\ref{iia3}) and that $A=\mathrm{const}$, one finds
\bea
\label{iiaaux1}
4S\langle\Lambda, \Gamma_i \Lambda\rangle+\langle\Lambda, \sgF_i \Gamma_{11}\Lambda\rangle +{1\over3} \langle \Lambda, \sgH_i\Gamma_{11}\Lambda\rangle=0~.
\eea
In what follows, we shall investigate the standard and massive IIA supergravities separately.

\subsubsection{Standard IIA supergravity with $S=0$}

In the case for which $S=0$, the dilatino KSEs ({\ref{kseadsk2iia}}) imply
that
\bea
\langle \Lambda, \slashed{G} \Gamma_{11} \Lambda \rangle =0~,
\eea
or equivalently, $W \wedge G=0$. As the $W$ span the tangent space
of $M^5$, it follows that $G=0$. Then, using the dilatino KSE
({\ref{kseadsk2iia}}) to eliminate the $F$ terms from ({\ref{iiaaux1}}),
we obtain
\bea
\langle \Lambda,  \sgH_i \Gamma_{11} \Lambda \rangle =0~,
\eea
which implies that $W \wedge H=0$. As the $W$ span the tangent
space of $M^5$, it follows that $H=0$ also.
The dilaton field equation (\ref{iiafeq}) then implies that
$F=0$ as well. However, for $S=0$, $G=0$, $H=0$ and $F=0$, the
the warp factor
field equation (\ref{warpfeq}) becomes inconsistent,
and so there are no $AdS_5$ solutions in standard IIA supergravity that preserve 24 supersymmetries.

\subsubsection{Massive IIA supergravity with $S \neq 0$}

On writing $G=\star_5 X$, where $X$ is a 1-form on $M^5$,
 the condition
 \bea
\label{snonzero1}
{5 \over 4}S \langle \Lambda, \Gamma_{11} \Lambda \rangle
+{1 \over 96} \langle \Lambda, \slashed{G} \Gamma_{11} \Lambda \rangle=0~,
\eea
which is derived from the dilatino KSE ({\ref{kseadsk2iia}}),
 can
be rewritten as
\bea
\label{snonzero2}
{5 \over 4}S \langle \Lambda, \Gamma_{11} \Lambda \rangle - {1 \over 4}
A^{-1} i_W X =0~.
\eea

Furthermore, the $G$ field equation implies that $dX=0$, and we assume\footnote{The invariance of $G$ under the vector fields constructed as Killing spinor bilinears
has not been proven for  massive IIA in complete generality, but it is expected to hold.} that ${\cal{L}}_W G=0$ which implies ${\cal{L}}_W X=0$. This condition, together with $dX=0$, gives that $i_W X$ is constant. Then it follows from ({\ref{snonzero2}}) that
$\langle \Lambda, \Gamma_{11} \Lambda \rangle$ is also constant.

On differentiating the condition $\langle \Lambda, \Gamma_{11} \Lambda \rangle=\mathrm{const}$ using
the gravitino KSEs, we obtain the condition
\bea
-{1 \over 4} F_{ij} \langle \Lambda, \Gamma^j \Lambda \rangle
+{1 \over 24} \langle \Lambda, \Gamma_{11} \slashed{G}_i  \Lambda \rangle =0~,
\eea
and hence
\bea
X^i F_{ij} \langle \Lambda, \Gamma^j \Lambda \rangle=0~.
\eea
However, using an argument directly analogous to that used
to show that the vector fields $W$ span the tangent space
of $M^5$, it follows that the vectors $\langle \Lambda, \Gamma^j \Lambda \rangle\partial_j$ also span the tangent space of $M^5$, see appendix B. Therefore,
\bea
i_X F=0~.
\eea
Next, act on the right-hand-side of the dilatino equation ({\ref{kseadsk2iia}})
with ${\slashed{X}} \Gamma_{11}$ and take the inner product with $\Lambda$.
On making use of $i_X F=0$, we find the condition
\bea
\langle \Lambda , X_{\ell_1} H_{\ell_2 \ell_3 \ell_4}
\Gamma^{\ell_1 \ell_2 \ell_3 \ell_4} \Lambda \rangle =0~,
\eea
and hence
\bea
\langle \Lambda, \Gamma_{11} \Gamma_{xyz} \Gamma_q \Lambda \rangle
\epsilon^{q \ell_1 \ell_2 \ell_3 \ell_4} X_{\ell_1} H_{\ell_2 \ell_3 \ell_4}=0~.
\eea
Again, as the vectors $\langle \Lambda, \Gamma_{11} \Gamma_{xyz} \Gamma^j \Lambda \rangle\partial_j$ span the tangent space of $M^5$, this condition implies that
\bea
X \wedge H=0~.
\eea
Another useful condition is to note that
${\cal{L}}_W X=0$ implies that
\bea
{\cal{L}}_W (D^i X_i)=0~,
\eea
and as the $W$ span the tangent space of $M^5$, it follows that
$D^i X_i$ must be constant. However the integral
of $D^i X_i$ over $M^5$ vanishes, and hence it follows that
\bea
D^i X_i=0~,
\eea
ie $X$ is co-closed. As it is also closed, $X$ and so $G$ are harmonic.
This condition, together with $dX=0$, imply that one can write
\bea
D^2 X^2 = 2 D^i X^j D_i X_j +2 X^j (D_i D_j
- D_j D_i) X^i = 2 D^i X^j D_i X_j +2 X^i X^j R^{(5)}_{i j}~.
\eea
On using the Einstein equation ({\ref{ein5}}), together with
the conditions $i_X F=0$, $X \wedge H=0$, we find
\bea
D^2 X^2 = 2 D^i X^j D_i X_j +
X^2 \big(-{1 \over 48}G^2 -{1 \over 2} S^2 -{1 \over 4} F^2 +{1 \over 6}H^2\big)~,
\eea
which can be written as
\bea
D^2 X^2 = 2 D^i X^j D_i X_j +
X^2 \big(2S^2+{3 \over 2}F^2\big)~,
\eea
on using the dilaton equation ({\ref{iiafeq}})
to eliminate the $G^2$ term. As the
right-hand-side of this expression is a sum of non-negative terms, an application of the
maximum principle implies that $X^2$ is constant{\footnote{The condition
$X^2=\mathrm{const}$ also follows from ${\cal{L}}_W X^2=0$ together with homogeneity.}} and
\bea
X^2 S^2=0~.
\eea
As $S \neq 0$, it follows that $X^2=0$, and hence $G=0$. Then
({\ref{snonzero1}}) implies that
\bea
\langle \Lambda, \Gamma_{11} \Lambda \rangle =0~,
\label{ass}
\eea
for all Killing spinors $\Lambda$.  However, this is a contradiction.

To see this, let the 12-dimensional vector space spanned by the Killing spinors $\Lambda$  be denoted by
$K$. Then the above condition implies that
\bea
\label{orthk}
\langle \Lambda_1, \Gamma_{11} \Lambda_2 \rangle =0~,
\eea
for all $\Lambda_1, \Lambda_2 \in K$. Denoting
\bea
\Gamma_{11} K = \{ \Gamma_{11} \Lambda : \Lambda \in K \}~,
\eea
the condition ({\ref{orthk}}) implies that
$\Gamma_{11}K \subseteq K^\perp$,
where
\bea
K^\perp = \{ \Psi: \langle \Psi, \Lambda \rangle =0\ {\rm for \ all \ }
\Lambda \in K \}~.
\eea
The dimension of space of all Majorana $Spin(9,1)$ spinors $\zeta$ satisfying the lightcone projection $\Gamma_+ \zeta=0$ is 16. As $K$ has dimension 12, $K^\perp$ has dimension 4.
As $\Gamma_{11} K$ is  12-dimensional it cannot be included in   $ K^\perp$ as required by the assumption (\ref{ass}).  Therefore there are no $AdS_5$
solutions in massive IIA supergravity which preserve 24 supersymmetries.

We would like to remark that the proof of this result is considerably simpler if $M^5$ is simply connected. As has already been proven, $G$ is harmonic.  On a simply connected $M^5$,  $G$ vanishes.  In such a case, (\ref{snonzero1}) again implies (\ref{ass}). Then the non-existence of such $AdS_5$ backgrounds follows from the argument produced above that
 (\ref{ass}) cannot hold for all Killing spinors.

\newsection{ $AdS_5 \times _w M^5$ solutions in IIB supergravity}

The active fields of $AdS_5\times_w M^5$ IIB backgrounds as well as the relevant field and KSEs have been determined in
\cite{iibads}.  In particular, in the conventions of \cite{iibads}, the metric and other form field strengths  are
\begin{eqnarray}
&&ds^2 = 2 du (dr + r h )+ A^2 (dz^2+ e^{2z/\ell} \sum_{a=1}^2 (dx^a)^2) +ds^2(M^5)~,
\cr
&&G=H,~~ P = \xi, ~~ F= Y \bigg( A^3 e^{2z \over \ell} du \wedge (dr+rh)
\wedge dz \wedge dx \wedge dy -d {\rm vol}(M^5) \bigg), \ \
\label{iibadsn}
\end{eqnarray}
where again we have written the background as a near-horizon geometry \cite{adshor},
with
\bea
h = -{2 \over \ell}dz -2 A^{-1} dA~,
\eea
 $A$ is the warp factor which is a smooth function on $M^5$, $G$ is the complex 3-form, $P$ encodes the (complexified) axion/dilaton gradients,
 $F$ is the real self-dual 5-form and $Y$ is a real scalar. The $AdS_5$ coordinates are $(u,r,z, x^a)$ and we introduce the frame
 $(\bbe^+, \bbe^-, \bbe^z, \bbe^a)$ as in (\ref{orthfr}).

For the analysis that follows, we shall use the  Bianchi identities
\bea
\label{bian}
d(A^5Y)=0, \qquad dH = iQ \wedge H - \xi \wedge {\bar{H}}
\eea
and the 10-dimensional Einstein equation along $AdS_5$ which gives the field equation
\bea
\label{einlpiib}
A^{-1} \nabla^2 A = 4Y^2 +{1 \over 48} \parallel H \parallel^2 -{4 \over \ell^2} A^{-2}
-4A^{-2} (dA)^2 \ ,
\eea
for the warp factor $A$.
The remaining Bianchi identities and bosonic field equations, which
are not necessary for the investigation  of $N=24$ solutions, can be found in \cite{iibads}.
 We also assume the same
regularity assumptions as for the eleven dimensional solutions, and remark that
({\ref{einlpiib}}) implies that $A$ is nowhere vanishing on $M^5$.

\subsection{The Killing spinors}

Solving the KSEs of IIB supergravity for $AdS_5\times_w M^5$ backgrounds along $AdS_5$, one finds that
the Killing spinors can be written as in (\ref{kkk}),  where now $\sigma_\pm$ and $\tau_\pm$ are Weyl
$Spin(9,1)$ spinors which depend only on the coordinates of $M^5$ that obey in addition the lightcone projections $\Gamma_\pm\sigma_\pm=\Gamma_\pm\tau_\pm=0$.

The remaining  independent KSEs are the gravitino parallel transport equations
\bea
\label{kseadsk1iib}
D_i^{(\pm)} \sigma_\pm = 0, \qquad D_i^{(\pm)} \tau_\pm = 0~,
\eea
where
\bea
D_i^{(\pm)} = D_i \pm {1 \over 2} \partial_i \log A -{i \over 2} Q_i
\pm {i \over 2} Y \Gamma_i \Gamma_{xyz}+\bigg(-{1 \over 96} \sgH_i +{3 \over 32}
{\slashed{H}}_i \bigg) C* \ ,
\eea
together with the dilatino KSEs
\bea
\label{iibalg1}
\bigg({1 \over 24} \slashed{H} + \slashed{\xi} C* \bigg) \sigma_\pm=0,
\qquad \bigg({1 \over 24} \slashed{H} + \slashed{\xi} C* \bigg) \tau_\pm =0~,
\eea
and some additional algebraic conditions which arise from the integration of the KSEs along the  $AdS_5$ subspace
\bea
\label{iibalg2}
\Xi^{(\pm)} \sigma_\pm=0, \qquad \bigg(\Xi^{(\pm)} \pm \ell^{-1} \bigg) \tau_\pm =0~,
\eea
where
\bea
\Xi^{(\pm)} = \mp {1 \over 2 \ell} -{1 \over 2} \Gamma_z \slashed{\partial} A \pm
{i \over 2} AY \Gamma_{xy} +{1 \over 96} A \Gamma_z \slashed{H} C* \ ,
\eea
and $C$ is the charge conjugation matrix.
Again, we have not made any assumptions on the form of the Killing spinors.

The counting of the Killing spinors, and the way in which one can construct
the $\sigma_\pm$, $\tau_\pm$ spinors from each other proceeds in exactly in the
same way as for the $D=11$ $AdS_5$ solutions.
So, again, for a generic $AdS_5 \times_w M^5$ solution, all of the Killing
spinors are generated by the $\sigma_+$ spinors, each of which
gives rise to 8 linearly independent spinors. The solutions therefore preserve $8k$ supersymmetries, where $k$ is equal
to the number of $\sigma_+$ spinors.

\subsection{$N=24$ $AdS_5$ solutions in IIB}

To proceed with the analysis first note that as a consequence of the homogeneity conjecture
proven in  \cite{homogen} is that the  solutions with 24 supersymmetries must
be locally homogeneous, with
\bea
\xi=0~.
\eea
Then, we set
\bea
 \Lambda = \sigma_+ + \tau_+~,
\eea
and define
\bea
\label{modisoiib}
W_i = A \langle \Lambda, \Gamma_{zxy} \Gamma_i \Lambda \rangle \ .
\eea
Then ({\ref{kseadsk1iib}}) implies that
\bea
D_{(i} W_{j)}=0~,
\eea
so $W$ is an isometry of $M^5$. Next, using ({\ref{kseadsk1iib}}),
we find
\bea
\label{iibaux1}
D_i \parallel \Lambda \parallel^2 = -\parallel \Lambda \parallel^2 A^{-1} D_i A
+{1 \over 48} {\rm Re} \langle \Lambda, \sgH_i C* \Lambda \rangle \ .
\eea
Furthermore, the algebraic condition ({\ref{iibalg2}}) implies
that
\bea
{1 \over 48} \slashed{H} C* \Lambda = \big(A^{-1} \Gamma^j D_j A -iY \Gamma_{xyz} \big) \Lambda
+\ell^{-1} A^{-1} \Gamma_z \big(\sigma_+-\tau_+\big) \ .
\eea
On substituting this condition back into ({\ref{iibaux1}}) we find
\bea
\label{iibaux2}
D_i \parallel \Lambda \parallel^2 = 2 \ell^{-1} A^{-1} {\rm Re} \langle \tau_+, \Gamma_i \Gamma_z \sigma_+ \rangle \ .
\eea
However, ({\ref{kseadsk1iib}}) also implies that
\bea
\label{iibaux3}
D^i \bigg( A {\rm Re} \langle \tau_+, \Gamma_i \Gamma_z \sigma_+ \rangle \bigg) =0~.
\eea
So combining this condition with ({\ref{iibaux2}}), we find
\bea
D^i D_i  \parallel \Lambda \parallel^2 +2 A^{-1} D^i A D_i  \parallel \Lambda \parallel^2 =0 \ .
\eea
A maximum principle argument then implies that $ \parallel \Lambda \parallel^2 $ is constant.
Then ({\ref{iibaux1}}) and ({\ref{iibaux2}}) imply
\bea
\label{iibconst1}
-\parallel \Lambda \parallel^2 A^{-1} D_i A
+{1 \over 48} {\rm Re} \langle \Lambda, \sgH_i C* \Lambda \rangle=0~,
\eea
or, equivalently
\bea
\label{iibconst2}
{\rm Re} \langle \tau_+, \Gamma_i \Gamma_z \sigma_+ \rangle =0 \ .
\eea

Next, we shall show that the spinors $\sigma_+$, $\tau_+$ are orthogonal with respect to
the inner product ${\rm Re} <,>$. To see this, note that ({\ref{iibalg2}}) implies that
\bea
\langle \tau_+, \Xi^{(+)} \sigma_+ \rangle =0, \qquad \langle \sigma_+, \big( \Xi^{(+)}+ \ell^{-1} \big) \tau_+ \rangle =0 \ .
\eea
On expanding out, and subtracting these two identities, one finds
that the real and imaginary parts of the resulting expression imply
\bea
\label{iibcc1}
\ell^{-1} {\rm Re} \langle \tau_+, \sigma_+ \rangle
+{\rm Re} \langle \tau_+, \Gamma_z \Gamma^i D_i A \sigma_+ \rangle =0~,
\eea
and
\bea
\label{iibcc2}
Y {\rm Re} \langle \tau_+, \Gamma_{xy} \sigma_+ \rangle +{1 \over 48}
{\rm Im} \langle \tau_+, \Gamma_z \slashed{H} C* \sigma_+ \rangle =0~,
\eea
respectively. On substituting ({\ref{iibconst2}}) into ({\ref{iibcc1}}), we find
that
\bea
{\rm Re} \langle \tau_+, \sigma_+ \rangle =0 \ .
\eea
For $N=24$ solutions there are 6 linearly independent $\sigma_+$ spinors,
and 6 linearly independent $\tau_+$ spinors, hence the spinors of the
type $\Lambda=\sigma_+ + \tau_+$ span a 12 dimensional vector space
over $\bR$, which we shall denote by $K$.

It is also particularly useful to note that the algebraic condition
({\ref{iibalg2}}) implies
\bea
&&{1 \over 2 \ell} \langle \Lambda, \Gamma_{xy}(\tau_+-\sigma_+)\rangle -{1 \over 2}
\langle \Lambda, \Gamma_{xyz} \Gamma^i D_i A \Lambda  \rangle
\cr &&~~~~~~~~~~~~~~
-{i \over 2}AY \parallel \Lambda \parallel^2
+{A \over 96} \langle \Lambda , \Gamma_{xyz} \slashed{H} C* \Lambda \rangle =0 \ .
\eea
On taking the real part of this expression, one finds
\bea
\label{liediib1}
W^i D_i A =0~,
\eea
where we have used the identity $\langle \Lambda, \Gamma_{xyz} \Gamma_{ijk} C* \Lambda \rangle =0$.

The condition ({\ref{liediib1}}) implies that
\bea
dA=0 \ .
\eea
This is because, by a straightforward  adaptation  of the analysis in
\cite{homogen}, it follows that the isometries $W$ generated by
the spinors $\Lambda \in K$ span the
tangent space of $M^5$, see also appendix B. So $A$ is constant, and the condition
({\ref{iibconst1}}) implies that
\bea
\label{iibconst1bb}
{\rm Re} \langle \Lambda, \sgH_i C* \Lambda \rangle=0~.
\eea
To proceed further, take the divergence of this expression. On making use of the
Bianchi identity for $H$ given in ({\ref{bian}}), together with
the KSE ({\ref{kseadsk1iib}}), we find the following condition:
\bea
\label{hhb1}
{\rm Re} \langle \Lambda, \bigg({9 \over 8} H_{\ell_1 \ell_2 i}
{\bar{H}}_{\ell_3 \ell_4}{}^i \Gamma^{\ell_1 \ell_2 \ell_3 \ell_4}
-{3 \over 4} H_{\ell_1 mn} {\bar{H}}_{\ell_2}{}^{mn} \Gamma^{\ell_1 \ell_2}
+{1 \over 4} H_{\ell_1 \ell_2 \ell_3} {\bar{H}}^{\ell_1 \ell_2 \ell_3}
\bigg) \Lambda \rangle =0 \ ,
\eea
where $\bar H$ is the complex conjugate of $H$.
Furthermore, the algebraic condition ({\ref{iibalg1}}) implies that
\bea
{\rm Re} \langle \Lambda, {1 \over 24} {\slashed{\bar{H}}} \slashed{H} \Lambda \rangle =0 \ .
\eea
On expanding this expression out, and adding it to ({\ref{hhb1}}), one
obtains the condition
\bea
{\rm Re} \langle \Lambda, H_{\ell_1 \ell_2 i}
{\bar{H}}_{\ell_3 \ell_4}{}^i \Gamma^{\ell_1 \ell_2 \ell_3 \ell_4} \Lambda \rangle =0~,
\eea
or equivalently
\bea
W^i \epsilon_i{}^{\ell_1 \ell_2 \ell_3 \ell_4} H_{\ell_1 \ell_2 j}
{\bar{H}}_{\ell_3 \ell_4}{}^j =0~.
\eea
Again, as the $W$ isometries span the tangent space of $M^5$, one obtains
\bea
\label{hhb2}
H_{[\ell_1 \ell_2 |i|}
{\bar{H}}_{\ell_3 \ell_4]}{}^i =0~.
\eea
Furthermore, on substituting this condition back into
\bea
\langle C* \Lambda, {\slashed{\bar{H}}} \slashed{H} \Lambda \rangle =0~,
\eea
which follows from ({\ref{iibalg1}}), we find
\bea
\label{iibgcon}
\langle C* \Lambda , \Lambda \rangle \parallel H \parallel^2 =0~.
\eea
So either $H=0$, or $\langle C* \Lambda , \Lambda \rangle=0$ for all $\Lambda \in K$.  We shall prove that $\langle C* \Lambda , \Lambda \rangle=0$ cannot
be satisfied for all $\Lambda$.

Indeed, suppose  that $\langle C* \Lambda , \Lambda \rangle=0$ for all
$\Lambda \in K$. We remark that $\langle C* \Lambda_1, \Lambda_2 \rangle$
is symmetric in $\Lambda_1, \Lambda_2$, and so $\langle C* \Lambda , \Lambda \rangle=0$ for all
$\Lambda \in K$ implies that
\bea
\label{ortho1}
\langle C* \Lambda_1, \Lambda_2 \rangle =0~,
\eea
for all $\Lambda_1,  \Lambda_2 \in K$. If we define
\bea
{\bar{K}} = \{ C* \Lambda : \Lambda \in K \}, \qquad K^\perp = \{\Psi : {\rm Re} \langle \Psi, \Lambda \rangle =0 \ {\rm for \ all \ } \Lambda \in K \}~,
\eea
then the condition ({\ref{ortho1}}) implies that ${\bar{K}} \subset K^\perp$.
However, this is not possible, because ${\bar{K}}$ is 12 dimensional, whereas
$K^\perp$ is 4-dimensional. So, one cannot have $\langle C* \Lambda , \Lambda \rangle=0$ for all
$\Lambda \in K$.

It follows that
\bea
H=0
\eea
and hence the spinors $\Lambda$ satisfy
\bea
\label{simpkse}
D_i \Lambda = \bigg({i \over 2} Q_i -{i \over 2} Y \Gamma_i \Gamma_{xyz} \bigg) \Lambda~,
\eea
for constant $Y$, $Y \neq 0$, with
\bea
Y^2 = {1 \over \ell^2 A^2}~,
\eea
as a consequence of ({\ref{einlpiib}}). The integrability condition of
({\ref{simpkse}}) implies that
\bea
\label{simpint1}
\bigg( R_{ijmn} - Y^2 (\delta_{im}\delta_{jn}-\delta_{in} \delta_{jm}) \bigg)
\Gamma^{mn} \Lambda=0~,
\eea
where we have used the Bianchi identity $dQ=0$. Then ({\ref{simpint1}})
gives that
\bea
{\rm Re} \langle \Lambda, \Gamma_{xyz} \bigg( R_{ijmn} - Y^2 (\delta_{im}\delta_{jn}-\delta_{in} \delta_{jm}) \bigg) \Gamma^n \Lambda \rangle =0~,
\eea
or equivalently
\bea
W^n \bigg( R_{ijmn} - Y^2 (\delta_{im}\delta_{jn}-\delta_{in} \delta_{jm}) \bigg)=0 \ .
\eea
As the isometries $W$ span the tangent space of $M^5$, it follows that
\bea
R_{ijmn} = Y^2 (\delta_{im}\delta_{jn}-\delta_{in} \delta_{jm})~,
\eea
and hence $M^5$ is locally isometric to the round $S^5$.

It follows that all (sufficiently regular) $AdS_5$ solutions with $N=24$
supersymmetries are locally isometric to $AdS_5 \times S^5$, with constant
axion and dilaton, and $G=0$.  This establishes that there are no distinct local geometries
for IIB $AdS_5\times M^5$ backgrounds that preserve strictly 24 supersymmetries.

\section{Concluding remarks}

We have proven, under some  assumptions,  a non-existence theorem for  $AdS_5\times_w M^{D-5}$, $D=10, 11$,  backgrounds that preserve strictly 24 supersymmetries in all 10- and 11-dimensional supergravity theories.
In particular we have demonstrated that such backgrounds cannot exist in 11-dimensional and (massive) IIA supergravities, and all such IIB backgrounds must be locally
isometric to the maximally supersymmetric $AdS_5\times S^5$ solution of the theory.

Our assumptions are that the fields must be smooth and the internal space $M^{D-5}$, $D=6,5$, must be  connected, compact and without boundary. Alternatively,
these assumptions can be summarized by saying that the data are such that the maximum principle applies. It turns out that these assumptions are required
to establish our results. It is known that if the compactness assumption for $M^6$ is removed, then the maximally supersymmetric $AdS_7\times S^4$ solution
 of 11-dimensional supergravity can be written locally as a warped $AdS_5\times_w M^6$ solution. This would appear to be a contradiction to our result for eleven
 dimensions, but for such a solution $M^6$ is not compact \cite{gutpap}.  Because of this, it is not apparent that the smoothness and global assumptions on $M^{D-5}$ can be removed.
 This in particular leaves open the possibility that there are $AdS_5\times_w M^{D-5}$ backgrounds in 10- and 11-dimensional supergravities but such
 backgrounds would either be singular or $M^{D-5}$ will not be compact and without boundary.  Another possibility of constructing $AdS_5$ backgrounds in IIB
with 24 supersymmetries is to take appropriate orbifolds of the maximally supersymmetric $AdS_5\times S^5$ solution of the theory. Though such a possibility
cannot be ruled out, it is unlikely.  It is also supported by the results of \cite{aharony2}, that there are no relevant ${\cal N}=3$ deformations
 of ${\cal N}=4$ theory.

The existence of a smooth $AdS_5$ background with compact without boundary  internal space in a 10- or 11-dimensional supergravity theory with distinct local geometry from that of maximally supersymmetric backgrounds would have raised the expectation that it should have been the AdS/CFT dual
to a ${\cal N}=3$  4-dimensional superconformal theory.  This would have been  in parallel with the well known duality that string theory on $AdS_5\times S^5$  is AdS/CFT dual
to ${\cal N}=4$ $U(N)$ gauge theory. Because both ${\cal N}=3$ and ${\cal N}=4$ theories have the same classical action, it is believed that in  perturbation
theory the two theories are indistinguishable. Though such a proof is not known, it may be  possible to demonstrate this by proving that quantum mechanically
  ${\cal N}=3$  Ward identities imply, using for example techniques similar to \cite{white},  that the symmetry enhances to ${\cal N}=4$.
In any case assuming that perturbatively the two theories cannot be distinguished, the possibility that remains is that if a theory exists with strictly
 ${\cal N}=3$ superconformal symmetry, it has to be intrinsically  non-perturbative. The properties of such a theory have been investigated in \cite{aharony2} and
 an F-theory construction has been proposed in \cite{garcia}.  Our results prove that the gravitational duals of strictly ${\cal N}=3$ superconformal
 theories cannot be smooth with compact without boundary internal spaces.  This is unlike the gravitational duals that preserve more than 16 supersymmetries  of other superconformal theories. The quantum mechanical consistency of ${\cal N}=3$ superconformal theories requires further investigation.

\vskip 0.5cm
\noindent{\bf Acknowledgements} \vskip 0.1cm
\noindent  We like to thank O. Aharony and A. Tomasiello for many helpful discussions and for asking questions about these $AdS_5$ backgrounds,  and thus encouraging us to publish these results; some of which
had been derived sometime ago. We also thank P. Howe  for his insightful remarks on aspects of ${\cal N}=3$ superconformal gauge theories, and J. Figueroa O'Farrill and A. Tseytlin for correspondence. JG is supported by the STFC grant, ST/1004874/1. GP is partially supported by the  STFC rolling grant ST/J002798/1.
\vskip 0.5cm

\setcounter{section}{0}\setcounter{equation}{0}

\appendix{Notation and conventions}

Our form conventions are as follows. Let $\omega$ be a k-form, then
\bea
\omega={1\over k!} \omega_{i_1\dots i_k} dx^{i_1}\wedge\dots \wedge dx^{i_k}~,~~~\omega^2_{ij}= \omega_{i\ell_1\dots \ell_{k-1}} \omega_{j}{}^{\ell_1\dots \ell_{k-1}}~,~~~
\omega^2= \omega_{i_1\dots i_k} \omega^{i_1\dots i_k}~.
\eea
We also define
\bea
{\slashed\omega}=\omega_{i_1\dots i_k} \Gamma^{i_1\dots i_k}~, ~~{\slashed\omega}_{i_1}= \omega_{i_1 i_2 \dots i_k} \Gamma^{i_2\dots i_k}~,~~~\slashed{\gom}_{i_1}= \Gamma_{i_1}{}^{i_2\dots i_{k+1}} \omega_{i_2\dots i_{k+1}}~,
\eea
where the $\Gamma_i$ are the Dirac gamma matrices.

The inner product $\langle\cdot, \cdot\rangle$ we use on the space of spinors is that for which space-like gamma matrices are Hermitian while time-like gamma
matrices are anti-hermitian, ie the Dirac spin-invariant inner product is $\langle\Gamma_0\cdot, \cdot\rangle$.  For more details on our  conventions
see \cite{mads, iiaads, iibads}.

\appendix{Homogeneity of Internal Spaces}

In this Appendix, we prove that for the $N=24$ $AdS_5$ solutions
in eleven-dimensional supergravity, the isometries on $M^6$ generated by the $\Lambda$ spinors via
\bea
\label{ddiag}
W = \langle \Lambda, \Gamma^i \Gamma_{xyz} \Lambda \rangle\, \partial_i~,
\eea
span the tangent space of $M^6$. The proof for this is a
straightforward adaptation of a similar result used in  the proof of the homogeneity conjecture
 \cite{homogen}. To begin, let $K$ denote the 12-dimensional
vector space spanned by the Killing spinors $\Lambda$.

Define the map $\varphi : K \otimes K \rightarrow TM^6$
by
\bea
\varphi (\Lambda_1, \Lambda_2) = \langle \Lambda_1, \Gamma^i \Gamma_{xyz} \Lambda_2 \rangle \partial_i~.
\eea
As $\varphi(\Lambda_1, \Lambda_2)=\varphi(\Lambda_2, \Lambda_1)$,
it follows that the $W$ span $T(M^6)$ iff $\varphi$ is surjective.
However, $\varphi$ is surjective iff the only vector $V \in T(M^6)$
satisfying
\bea
V^i \varphi(\Lambda_1, \Lambda_2)_i =0~,
\eea
for all $\Lambda_1, \Lambda_2 \in K$ is $V=0$, i.e. the perpendicular complement of the image of $\varphi$ is trivial.

Suppose, for a contradiction, that the perpendicular complement
of the image of $\varphi$ is not trivial. Then there exists
nonzero $V \in T(M^6)$ such that
\bea
\label{noinclude}
V^i \Gamma_i \Gamma_{xyz} \Lambda \in K^\perp~,
\eea
for all $\Lambda \in K$, where
\bea
K^\perp = \{ \Psi : \langle \Psi, \Lambda \rangle =0 \ \ {\rm for \ all \ } \Lambda \in K \}~.
\eea
Observe that $K\oplus K^\perp$ is a 16-dimensional vector space spanned by the Majorana $Spin(10,1)$ spinors $\zeta$ that satisfy the lightcone projection $\Gamma_+\zeta=0$.
Thus as $K$ is 12-dimensional, $K^\perp$ is a 4-dimensional subspace.

As $V \neq 0$, the kernel of the map
$V^i \Gamma_i \Gamma_{xyz}:  K\rightarrow K^\perp$ is zero and so it is injective. However this is not possible as the image $V^i \Gamma_i \Gamma_{xyz}(K)$
is 12-dimensional while $K^\perp$ is 4-dimensional.  Thus the hypothesis that $V\not=0$ is not valid and $\varphi$ is surjective, and so the vectors $W$ span the
tangent space of $M^6$.

The argument for the $AdS_5$ backgrounds of massive IIA supergravity is the same as that described above upon replacing $M^6$ with $M^5$. It also generalizes  for the $AdS_5$ solutions in IIB
supergravity, after replacing the norm $<,>$ with  ${\rm Re} <,>$,
and $M^6$ with  $M^5$ throughout.

\end{document}